\documentclass[twocolumn,showpacs,preprintnumbers,amsmath,amssymb]{revtex4}
\usepackage{graphicx}
\usepackage{dcolumn}
\usepackage{bm}

\begin{document}
\preprint{APS/123-QED}

\title{Isospin effects on the mass dependence of balance energy}

\author{Sakshi Gautam$^{1}$ }
\author{Aman D. Sood$^{2}$}%
\email{amandsood@gmail.com} \affiliation{$^{1}$Department of
Physics, Panjab University, Chandigarh -160 014, India.}
\affiliation{$^{2}$SUBATECH, Laboratoire de Physique Subatomique et des Technologies Associ\'{e}es, Universit\'{e} de Nantes - IN2P3/CNRS - EMN \\
4 rue Alfred Kastler, F-44072 Nantes, France.}


\date{\today}

\begin{abstract}
We study the effect of isospin degree of freedom on balance energy
throughout the mass range between 50 and 350 for two sets of
isotopic systems with N/Z = 1.16 and 1.33 as well as isobaric
systems with N/Z = 1.0 and 1.4. Our findings indicate that
different values of balance energy for two isobaric systems may be
mainly due to the Coulomb repulsion. We also demonstrate clearly
the dominance of Coulomb repulsion over symmetry energy.
\end{abstract}

\pacs{25.70.Pq, 25.70.-z}

\maketitle

\section{Introduction}
The investigation of the system size effects in various phenomena
of heavy-ion collisions has attracted a lot of attention. For
example, in the low-energy regime where phenomena like fusion,
fission, cluster radioactivity, formation of super heavy nuclei,
etc. \cite{puri}, take place, the contribution of the Coulomb
force toward barrier has been reported to scale with the mass and
charge of the colliding nuclei \cite{jian06}. Similarly, the
system size dependences have been reported in various other
phenomena like particle production, multifragmentation, collective
flow (of nucleons/fragments), density, temperature and so on. For
instance, in Ref. \cite{stur01} the power law scaling
($\varpropto$ A$^{\tau}$) of pion/kaon production with the size of
the system has been reported. Similar power law behavior for the
system size dependence has been reported for the multiplicity of
various types of fragments also \cite{sing}. The collective
transverse in-plane flow which reflects the competition between
attractive and repulsive interactions has been investigated
extensively during the past three decades and has been found to
depend strongly on the combined mass of the system \cite{ogli} in
addition to the incident energy \cite{pan,zhang} as well as
colliding geometry \cite{zhang}. The energy dependence of
collective transverse in-plane flow has led us its disappearance.
It has now been well established that there exists a particular
incident energy at which the attractive and repulsive parts of
nuclear interactions counterbalance each other and the net flow
disappears. This energy has been termed as the balance energy
(E$_{bal}$) or the energy of vanishing flow (EVF)
\cite{bon87,krof}. E$_{bal}$ has been found to depend strongly on
the combined mass of the system. A power law mass dependence
($\varpropto$ A$^{\tau}$) of E$_{bal}$ also has been reported
\cite{mota,mag1,sood2,sood3}. Earlier power law parameter $\tau$
was supposed to be close to -1/3 (resulting from the interplay
between attractive mean field and repulsive nucleon-nucleon (nn)
collisions) \cite{mota}, whereas recent studies showed a deviation
from the above-mentioned power law \cite{sood1,mag1,sood2,sood3}
where $\tau$ was close to -0.45. Recently Sood and Puri
\cite{sood2}, reported the power law mass dependence $\varpropto$
1/$\sqrt{A}$ for heavier nuclei which suggested the increasing
importance of the Coulomb interactions. Another interesting study
of the mass dependence of density and temperature reveals that the
maximum temperature is insensitive toward the combined mass of the
system, whereas maximum density scales with the size of the system
\cite{khoa}. However, Sood and Puri \cite{sood4} recently
suggested the power law mass dependence for the temperature at
E$_{bal}$, which was later confirmed experimentally by Wang
\emph{et al.} \cite{wang}. In another study Sood and Puri
\cite{sood3} studied the effect of momentum-dependent interactions
(MDI) on the collective flow as well as its disappearance through
out the mass range (from $^{12}$\emph{C} + $^{12}$\emph{C} to
$^{197}$\emph{Au} + $^{197}$\emph{Au}). They found that the impact
of MDI differs in lighter nuclei as compared to the heavier ones.
\par
With the availability of high intensity radioactive beams at many
facilities \cite{rib1}, the effects of isospin degree of freedom
in nuclear reactions can be studied in more details over a wide
range of masses at different incident energies and colliding
geometries. Such studies can help to isolate the isospin dependent
part of the nuclear matter equation of state (EOS) which is vital
to understand the astrophysical phenomena such as neutron stars,
supernovae explosions, and so on. In our recent communication
\cite{gaum}, we have studied the E$_{bal}$ for \emph{$^{58}$Ni +
$^{58}$Ni} and \emph{$^{58}$Fe + $^{58}$Fe} data \cite{pak}. Our
calculations are able to reproduce the experimental data within
3\% on the average over all colliding geometries (guided by Ref.
\cite{pak}). The good agreement of our calculations with the data
motivated us to study the isospin effects on the E$_{bal}$
throughout the mass range. It is worth mentioning here that the
isospin-dependent quantum molecular dynamics (IQMD) model has also
been able to reproduce the other data (e.g. high-energy proton
spectra, gamma production) in the incident energies relevant in
this article \cite{ger98,gamm}. The
present aim is atleast twofold.\\
 (1) To study the effect of
isospin degree of freedom on the E$_{bal}$ throughout the
mass range.\\
 (2) As reported in literature, the isospin
dependence of collective flow has been explained as the
competition among various reaction mechanisms, such as
nucleon-nucleon collisions, symmetry energy, surface property of
the colliding nuclei, and Coulomb force. The relative importance
among these mechanisms is not yet clear \cite{li}. In the present
study, we aim to shed light on the relative importance among the
above-mentioned reaction mechanisms by taking pairs of isotopic as
well as isobaric systems throughout the mass range. Section II
describes the model in brief. Section III explains the results and
discussion and Sec. IV summarizes the results.
\section{The model}
The present study is carried out within the framework of IQMD
model \cite{hart,aichqmd,qmd2}. In the IQMD model, each nucleon
propagates under the influence of mutual two- and three-body
interactions. The propagation is governed by the classical
equations of motion:
\begin{equation}
\dot{{\bf r}}_i~=~\frac{\partial H}{\partial{\bf p}_i}; ~\dot{{\bf
p}}_i~=~-\frac{\partial H}{\partial{\bf r}_i},
\end{equation}
where H stands for the Hamiltonian which is given by:
\begin{eqnarray}
 H = \sum_i^{A} {\frac{{\bf p}_i^2}{2m_i}}~~~~~~~~~~~~~~~~~~~~~~~~~~~~~~~~~~~~~~~~~~~~~~~~\nonumber\\
 ~~~~~+ \sum_i^{A} ({V_i^{Sky} + V_i^{Yuk} + V_i^{Coul} +
V_i^{mdi} + V_i^{sym}}).
\end{eqnarray}
The $V_{i}^{Sky}$, $V_{i}^{Yuk}$, $V_{i}^{Coul}$, $V_i^{mdi}$, and
$V_i^{sym}$ are, respectively, the Skyrme, Yukawa, Coulomb,
momentum dependent interactions (MDI), and symmetry potentials.
The MDI are obtained by parameterizing the momentum dependence of
the real part of the optical potential. The final form of the
potential reads as \cite{aichqmd}
\begin{equation}
U^{mdi}\approx t_{4}ln^{2}[t_{5}({\bf p_{1}}-{\bf
p_{2}})^{2}+1]\delta({\bf r_{1}}-{\bf r_{2}}).
\end{equation}
Here $t_{4}$ = 1.57 MeV and $t_{5}$ = $5\times 10^{-4} MeV^{-2}$.
A parameterized form of the local plus MDI potential is given by
\begin{equation}
U=\alpha \left({\frac {\rho}{\rho_{0}}}\right) + \beta
\left({\frac {\rho}{\rho_{0}}}\right)^{\gamma}+ \delta
ln^{2}[\epsilon(\rho/\rho_{0})^{2/3}+1]\rho/\rho_{0}.
\end{equation}
The parameters $\alpha$, $\beta$, $\gamma$, $\delta$, and
$\epsilon$ are listed in Ref \cite{aichqmd}.
\par
\section{Results and discussion}
For the present study, we simulate the various reactions in the
incident energy range between 40 and 150 MeV/nucleon in small
steps of 10 MeV/nucleon. In particular, we simulate the reactions
$^{26}$Mg+$^{26}$Mg, $^{65}$Zn+$^{65}$Zn, $^{91}$Mo+$^{91}$Mo,
$^{117}$Xe+$^{117}$Xe, $^{164}$Os+$^{164}$Os having N/Z = 1.16 and
reactions $^{28}$Mg+$^{28}$Mg, $^{70}$Zn+$^{70}$Zn,
$^{98}$Mo+$^{98}$Mo, $^{126}$Xe+$^{126}$Xe, and
$^{177}$Os+$^{177}$Os having N/Z = 1.33, respectively, at
semicentral impact parameter range 0.35 - 0.45. The N/Z for a
given pair is varied by adding the neutron content only keeping
the charge fixed, so that the effect of the Coulomb potential is
same for a given mass pair. However this will lead to the increase
in mass for systems with higher neutron content. In Ref.
\cite{bon10}, there is evidence that N/Z is an order parameter. We
use a soft equation of state with MDI labeled as SMD. We also use
anisotropic standard isospin- and energy dependent nucleon-nucleon
cross section $\sigma$ = 0.8 $\sigma$$_{NN}$$^{free}$. The details
about the elastic and inelastic cross sections for proton-proton
and proton-neutron collisions can be found in Ref.
\cite{cug,hart}. The cross sections for neutron-neutron collisions
are assumed to be equal to the proton-proton cross sections. Some
studies even took constant cross section \cite{kum}. It is worth
mentioning that the results with the above choice of equation of
state and cross section were in good agreement with the data
\cite{gaum}. The choice of reduced cross section has also been
motivated by Ref. \cite{daff} as well as many previous studies
\cite{zhou94}. In Ref. \cite{gaum}, we found the effect of angular
distribution of scattering cross sections on E$_{bal}$ to be
negligible. Recently, Sood and Puri \cite{sood1} have discussed
the role of different cross sections on the E$_{bal}$ throughout
the mass range between 47 and 394. They found the effect of
different cross sections to be consistent throughout the mass
range. The largest cross section gives the more positive flow
(hence smaller E$_{bal}$) followed by the second largest cross
section. Similar results were obtained in Ref. \cite{gaum}. The
reactions are followed until the transverse flow saturates. The
saturation time varies form 100 fm/c for lighter masses to 300
fm/c for heavier masses. For the transverse flow, we use the
quantity "\textit{directed transverse momentum $\langle
p_{x}^{dir}\rangle$}" which is defined as
\cite{sood2,sood3,sood4,leh}
\begin {equation}
\langle{p_{x}^{dir}}\rangle = \frac{1} {A}\sum_{i=1}^{A}{sign\{
{y(i)}\} p_{x}(i)},
\end {equation}
where $y(i)$ is the rapidity and $p_{x}$(i) is the momentum of
$i^{th}$ particle. The rapidity is defined as
\begin {equation}
Y(i)= \frac{1}{2}\ln\frac{{\textbf{{E}}}(i)+{{\textbf{p}}}_{z}(i)}
{{{\textbf{E}}}(i)-{{\textbf{p}}}_{z}(i)},
\end {equation}
where ${\textbf{E}}(i)$ and ${\textbf{p}_{z}}(i)$ are,
respectively, the energy and longitudinal momentum of $i^{th}$
particle. In this definition, all the rapidity bins are taken into
account.
\par
\begin{figure}[!t]
\centering \vskip 0cm
\includegraphics[angle=0,width=8cm]{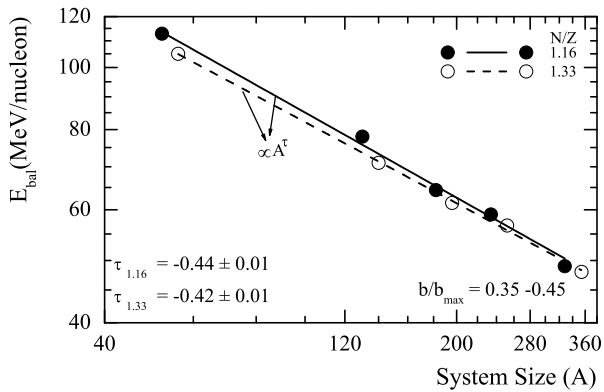}
\vskip 0cm \caption{E$_{bal}$ as a function of combined mass of
system. Various symbols and lines are explained in the
text.}\label{fig1}
\end{figure}
In Fig. 1, we display the E$_{bal}$ as a function of combined mass
of the system for two sets of isotopic systems with different
neutron content. Closed (open) circles represent systems with less
(more) neutron content. Lines are the power law fit $\varpropto$
A$^{\tau}$. As expected, both the sets of isotopic masses follow a
power law behavior $\varpropto$ A$^{\tau}$, where $\tau$ = -0.44
$\pm$ 0.01 for systems with less neutron content (labeled as
$\tau_{1.16}$) and -0.42 $\pm$ 0.01 for systems with more neutron
content (labeled as $\tau_{1.33}$). Note that the values of
$\tau_{1.16}$ and $\tau_{1.33}$ are very close to the previous
values of $\tau$ in Ref. \cite{mag1}. Interestingly, isospin
effects are not visible for any mass system. As reported in
literature \cite{pak,li}, system with more neutron content has
higher E$_{bal}$ which has been attributed mainly to the
above-mentioned reaction mechanisms and to the fact that
neutron-neutron or proton-proton cross section is a factor of 3
lower than the neutron-proton cross section. However, in our case
system size effects seem to dominate the isospin effects
throughout the mass range. It is worth mentioning that the effect
of the Coulomb force is almost the same for a given pair of
masses.
 \par
As a next step, we take the pairs of isobars with N/Z = 1.0 and
1.4.
 We simulate the reactions $^{24}$Mg+$^{24}$Mg,
$^{58}$Cu+$^{58}$Cu, $^{72}$Kr+$^{72}$Kr, $^{96}$Cd+$^{96}$Cd,
$^{120}$Nd+$^{120}$Nd, $^{135}$Ho+$^{135}$Ho, having N/Z = 1.0 and
reactions $^{24}$Ne+$^{24}$Ne, $^{58}$Cr+$^{58}$Cr,
$^{72}$Zn+$^{72}$Zn, $^{96}$Zr+$^{96}$Zr, $^{120}$Sn+$^{120}$Sn,
and $^{135}$Ba+$^{135}$Ba, having N/Z = 1.4, respectively. Here
N/Z for a given pair is changed by varying both the proton and
neutron content keeping the mass fixed.
\begin{figure}[!t]
\centering
 \vskip 0cm
\includegraphics[angle=0,width=8cm]{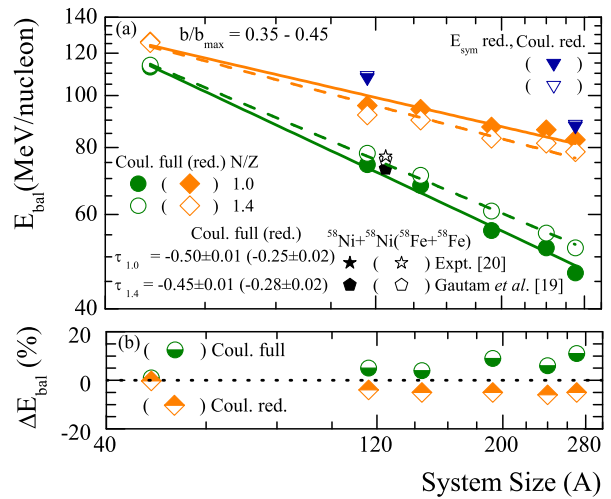}
 \vskip 0.3cm
\caption{(a) E$_{bal}$ as a function of combined mass of system.
(b) The percentage difference $\Delta E_{bal} (\%) $ as a function
of combined mass of system. Solid (open) symbols are for N/Z = 1.0
(1.4). Various symbols and lines are explained in the
text.}\label{fig2}
\end{figure}
In Fig. 2(a), we display the E$_{bal}$ as a function of combined
mass of the system for the two sets of isobars. We also display
the experimental data \cite{pak} and our corresponding theoretical
calculations \cite{gaum} (both displaced horizontally for clarity)
in the range of colliding geometry used in the present study.
Solid (open) stars represent the experimentally measured E$_{bal}$
for the reaction \emph{$^{58}$Ni + $^{58}$Ni} (\emph{$^{58}$Fe +
$^{58}$Fe}) having N/Z = 1.07 (1.23) in the impact parameter bin
0.28 $<$ b/b$_{max}$ $<$ 0.39 (guided by Ref. \cite{pak}).
Pentagons represent the corresponding results of our theoretical
calculations. Clearly our results are in good agreement with the
data. Our theoretical calculations are able to reproduce the
experimentally measured E$_{bal}$ for both the systems. Solid and
open green circles represent the E$_{bal}$ for systems with less
and more neutron content, respectively. Lines are power law fit
$\propto$ A$^{\tau}$. Interestingly, throughout the mass range, a
more neutron-rich system has a higher E$_{bal}$. The calculated
E$_{bal}$ fall on the line that is a fit of power law nature
($\propto$ A$^{\tau}$), where $\tau$ = -0.45 $\pm$ 0.01 and -0.50
$\pm$ 0.01 for N/Z = 1.4
 and 1.0 (labeled by $\tau_{1.4}$ and $\tau_{1.0}$), respectively. The different values of $\tau$ for two
curves can be attributed to the larger role of Coulomb force in
the case of systems with more proton (less neutron) content. Our
value of $\tau$$_{1.4}$ is equal/close to the value -0.45/-0.42 in
Ref. \cite{mag1} both of which show deviation from the standard
value $\simeq$ -1/3 \cite{mota} where analysis was done for
lighter mass nuclei only ($\leq$ 200). However, when heavier
systems like \emph{$^{139}$La + $^{139}$La} and \emph{$^{197}$Au
+} \emph{$^{197}$Au} were included, $\tau$ increased to -0.45
\cite{mag1}, suggesting the increasing importance of Coulomb
repulsion. A further analysis in Ref. \cite{sood2} showed that
when only heavier nuclei were taken into account, the value of
$\tau$ increased to -0.53 which is very close to our value of
$\tau_{1.0}$ (-0.50) in the present case. Although the mass range
in the present study and in Ref. \cite{sood2} are differ
substantially the present power law parameter $\tau_{1.0}$ is very
close to the power law parameter
 of Ref. \cite{sood2} which can be attributed to the extra Coulomb repulsion in case of
systems with more proton (less neutron) content as well as Ref.
\cite{sood2}. This indicates that the difference in the E$_{bal}$
for a given pair of isobaric systems may be dominantly due to the
Coulomb potential, which is further supported by the fact that
since both asymmetry energy and nucleon-nucleon cross section add
to the repulsive interactions, so both lead to the reduction of
E$_{bal}$. In systems with more neutron content, role of asymmetry
energy could be larger, whereas effects due to isospin-dependent
cross section could play a dominant role in systems with less
neutron content (more proton content). Therefore, there is a
possibility that the combined effect of asymmetry energy and cross
section could be approximately the same for two isobaric systems
with different neutron and proton content.
\par
To demonstrate the dominance of Coulomb, we have also calculated
the E$_{bal}$ throughout the mass range in the present study for
isobaric pairs with Coulomb being reduced by a factor of 100. The
results are displayed in Fig. 2(a) with solid and open (orange)
diamonds representing systems with less and more neutron content,
respectively. Lines represent power law fit $\propto$ A$^{\tau}$.
One can clearly see the dominance of Coulomb repulsion in both the
mass dependence as well as in isospin effects. The value of
$\tau_{1.4}$ and $\tau_{1.0}$ are now, respectively, -0.28 $\pm$
0.02 and -0.25 $\pm$ 0.02. There is large enhancement in the
E$_{bal}$ for medium (eg. $^{58}$Cu+$^{58}$Cu) and heavy
($^{120}$Nd+$^{120}$Nd) mass systems thus reducing the value of
both $\tau_{1.0}$ and $\tau_{1.4}$. The effect is small in lighter
masses ($^{24}$Mg+$^{24}$Mg). One should also note that when we
have full Coulomb included in our calculations, for medium and
heavy mass systems, E$_{bal}$ is less for systems having more
proton (less neutron) content. However, the trend is reversed when
we reduce the Coulomb. Now the systems with more neutron content
have less E$_{bal}$. This is because of the fact that the reduced
Coulomb repulsion leads to higher E$_{bal}$. As a result, the
density achieved during the course of the reaction will be more
due to which the impact of the repulsive symmetry energy will be
more in neutron-rich systems, which in turn leads to less
E$_{bal}$ for neutron-rich systems and hence to the opposite trend
for $\tau_{1.4}$ and $\tau_{1.0}$ for two different cases (Coulomb
full and reduced). To check this point we have also calculated the
E$_{bal}$ by reducing the strength of both symmetry energy as well
as Coulomb potential for isobaric pairs with combined mass of the
system = 116 and 270 (shown by blue triangles in Fig. 2(a)). We
find that both the systems of a given isobaric pair have same
E$_{bal}$. The above discussion clearly points towards the
dominance of Coulomb repulsion over asymmetry energy for medium as
well as heavy mass systems, whereas their impact is small in
lighter masses.
\begin{figure}[!t] \centering
 \vskip 0cm
\includegraphics[angle=0,width=8cm]{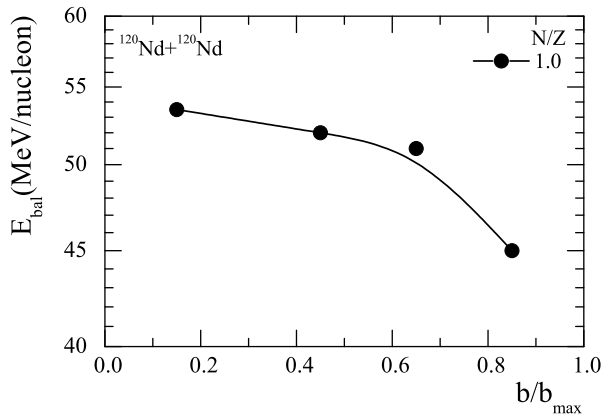}
 \vskip 0cm
\caption{(a) E$_{bal}$ as a function of impact parameter for
system having mass 240 with N/Z = 1.0.}\label{fig3}
\end{figure}
\par
In Fig. 2(b), we display the percentage difference $\triangle
E_{bal}$(\%) between the systems of isobaric pairs as a function
of combined mass of system where $\triangle E_{bal} (\%) =
\frac{E_{bal}^{1.4}-E_{bal}^{1.0}}{E_{bal}^{1.0}}\times 100$.
Superscripts to the E$_{bal}$ represent different N/Z. Half filled
(green) circles are for full Coulomb and half filled (orange)
diamonds are for reduced Coulomb. Negative (positive) values of
$\triangle E_{bal} (\%)$ shows that the E$_{bal}$$^{1.0}$ is more
(less) than E$_{bal}$$^{1.4}$. From Fig. 2(b) (circles), we see
that the percentage difference between the two masses of a given
pair is larger for heavier masses as compared to the lighter ones.
In lighter masses, the magnitude of Coulomb repulsion is small so
 there is small difference, whereas in heavier masses, due to the large magnitude of Coulomb repulsion, there is a large difference in the E$_{bal}$ for a given
 pair of isobars. However, this trend is not visible when we reduce the Coulomb
(diamonds). The values of $\triangle E_{bal} (\%)$ is almost
constant for medium and heavy masses.
 \par
 In Fig. 3, we display the E$_{bal}$ as a
function of impact parameter for the system
 having mass 240 with N/Z = 1.0. In the present case, the E$_{bal}$ seems to decrease with increase in
 impact parameter (in contrast to earlier studies where E$_{bal}$ increases with increase in impact parameter \cite{mag1,pak,gaum}),  since at a higher impact parameter there will be large transverse
 flow of nucleons due to the dominant Coulomb repulsion. This could also play a part for larger
 difference in the E$_{bal}$ at higher colliding geometries
 \cite{gaum,pak,li} for systems with different neutron and proton content.
It is worth mentioning that
 dominance of Coulomb repulsion in isospin effects has also been
 reported by Ref. \cite{daff}.
 \par
 \section{Summary}
 We have studied the effect of isospin degree of
 freedom on the E$_{bal}$ throughout the mass range for two sets of isotopic systems with N/Z = 1.16 and
 1.33
 as well as isobaric systems with N/Z = 1.0 and 1.4. Our results
 indicate that the difference between the E$_{bal}$ for the two isobaric
 systems may be mainly due to the Coulomb repulsion. We have also shown clearly the dominance of Coulomb repulsion over symmetry energy.
Our findings
 also point that the larger magnitude of isospin effects in
 E$_{bal}$ at peripheral collisions as compared to central collisions may be dominantly due to the Coulomb
 repulsion.
 \par
 We thank R. K. Puri, C. Hartnack and
J. Aichelin for useful suggestion and enlightening discussion. The
work is supported by Indo-French project no. 4104-1.

\end{document}